\documentclass[letterpaper, 10 pt, conference]{ieeeconf}  % Comment this line out if you need a4paper

\usepackage{amsfonts}
\usepackage{amsmath}
\usepackage{booktabs}
\usepackage{caption} 
\let\labelindent\relax
\usepackage{enumitem}
\usepackage{graphicx}
\usepackage{ifthen}
\graphicspath{ {./images/} }
\IEEEoverridecommandlockouts                              % This command is only needed if 
\title{\LARGE \bf
Goal-Directed Occupancy Prediction for Lane-Following Actors
}

\author{
  Poornima Kaniarasu$^{1,2}$, Galen Clark Haynes$^{1}$, and Micol Marchetti-Bowick$^{1}$
  \thanks{$^{1}$ Research conducted at Uber ATG, 50 33rd St, Pittsburgh, PA, USA. Email:
  {\tt\small gch@uber.com};
  {\tt\small mmarchettibowick@uber.com}}
  \thanks{$^{2}$ Previously affiliated with Uber ATG (Jan 11, 2016 - Dec 18, 2019). Email: {\tt\small mkporkodi@gmail.com}}
}

\begin{document}

\setlength{\abovedisplayskip}{3pt}
\setlength{\belowdisplayskip}{3pt}

\maketitle
\thispagestyle{empty}
\pagestyle{empty}

%%%%%%%%%%%%%%%%%%%%%%%%%%%%%%%%%%%%%%%%%%%%%%%%%%%%%%%%%%%%%%%%%%%%%%%%%%%%%%%%
\begin{abstract}

Predicting the possible future behaviors of vehicles that drive on shared roads is a crucial task for safe autonomous driving. Many existing approaches to this problem strive to distill all possible vehicle behaviors into a simplified set of high-level actions. However, these action categories do not suffice to describe the full range of maneuvers possible in the complex road networks we encounter in the real world. To combat this deficiency, we propose a new method that leverages the mapped road topology to reason over possible goals and predict the future spatial occupancy of dynamic road actors.  We show that our approach is able to accurately predict future occupancy that remains consistent with the mapped lane geometry and naturally captures multi-modality based on the local scene context while also not suffering from the mode collapse problem observed in prior work.

\end{abstract}

%%%%%%%%%%%%%%%%%%%%%%%%%%%%%%%%%%%%%%%%%%%%%%%%%%%%%%%%%%%%%%%%%%%%%%%%%%%%%%%%
\section{INTRODUCTION}

% Paragraph 1: introduction of autonomous vehicle hard problems and where spatial prediction fits into the bigger picture
In order for autonomous vehicles to successfully navigate a world filled with dynamic actors, they must perform a wide variety of challenging tasks. These include perceiving objects in a scene, forecasting actor motion, and determining an appropriate ego-trajectory to drive both safely and comfortably. While the domains of object detection and motion planning are well represented in the literature, the problem of predicting another actor's future motion is a newer but rapidly growing area of study that is critically important for autonomous driving~\cite{Rudenko2019}. % Other actors in busy city scene are driving along lanes and driveways, making navigational decisions, as well as interacting with other actors and traffic control elements.  Being able to predict these intentions and actions is thus critical to building a successful autonomous vehicle. 

The most challenging aspect of the prediction problem is that the future is inherently ambiguous. Consider a vehicle approaching an intersection. Although we can use the actor's current and past motion, along with contextual cues, to get an idea of where the actor will go, the driver could quite easily change their mind about which direction to travel while navigating the intersection. As a result, in order to successfully tackle the problem of long-term motion prediction (over a 5-10 second horizon), we must directly handle this ambiguity by considering multiple possible futures. 
%This implies that the ground truth is not a single trajectory but a multi-modal distribution over trajectories, even though we ultimately only observe one sample from that distribution. 

\begin{figure}[tbp]
\centering
\begin{minipage}[b]{\linewidth}
\centering
\includegraphics[width=0.47\linewidth]{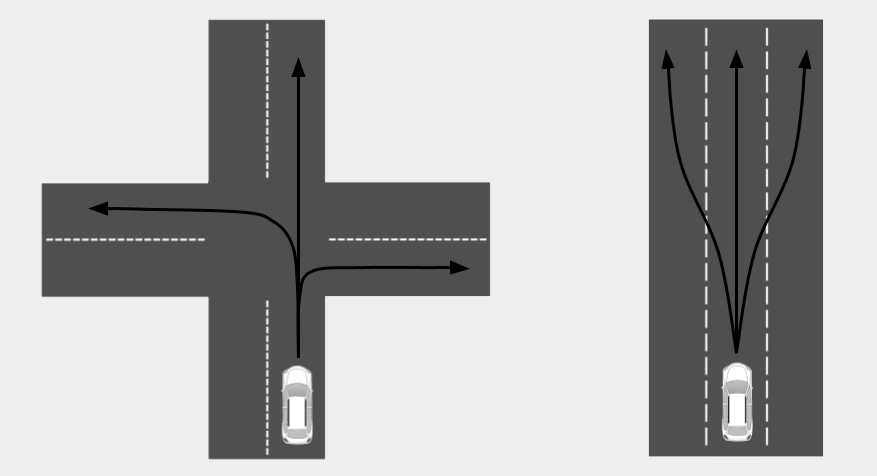}
\hspace{.1cm}
\includegraphics[width=0.48\linewidth]{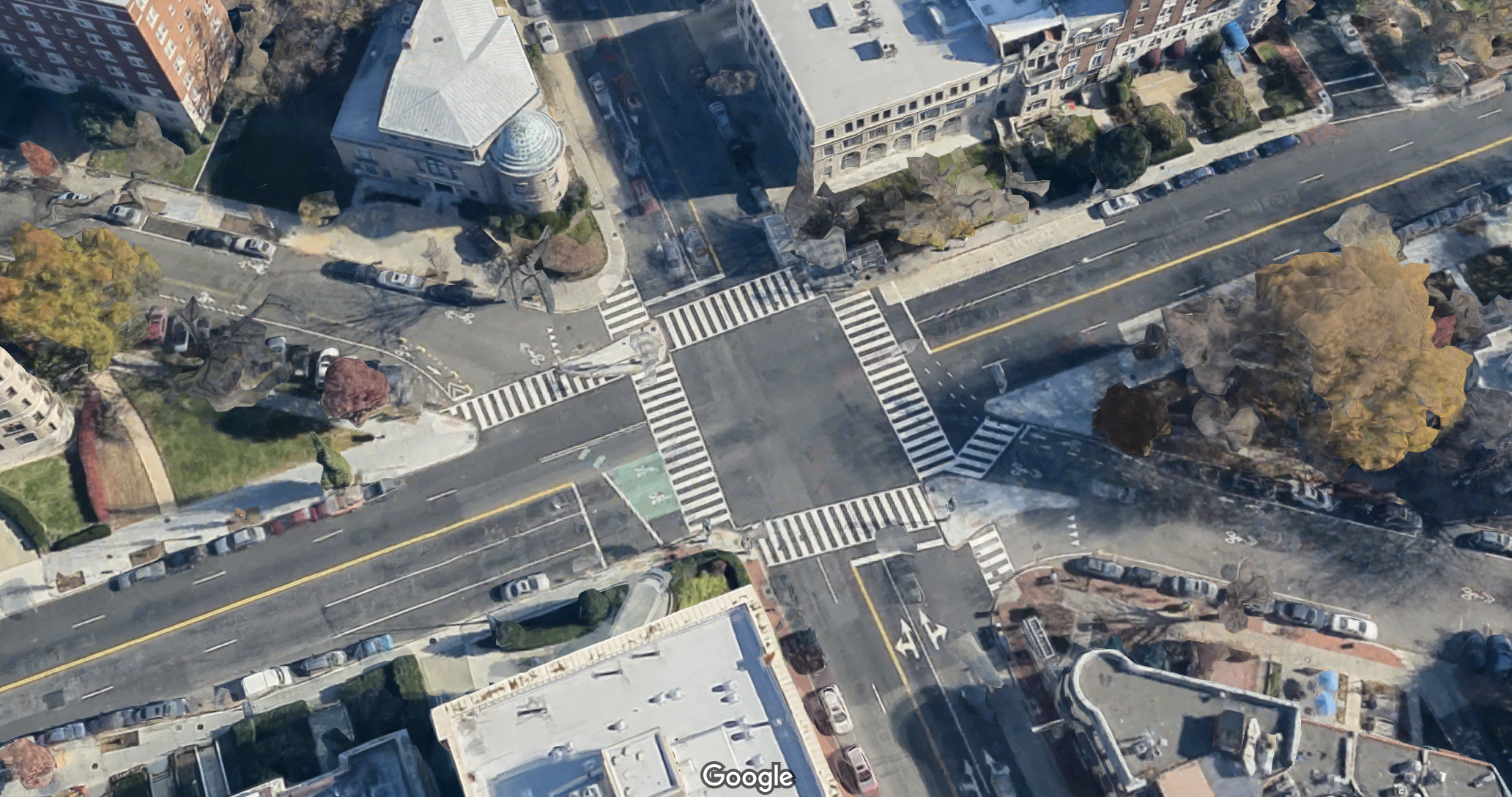}
\end{minipage}
\caption{(a) Existing multi-modal prediction models assume that actor behavior can be summarized by a set of high-level maneuvers such as left/right/straight, as illustrated on the left. (b) However, many real-world intersections are much more complex than this, such as the 6-way intersection in Washington, DC shown on the right (source: Google Maps).}
\label{fig:left-right-straight}
\end{figure}

One way to achieve this is to predict a distribution over the actor's future motion. Previous methods have primarily explored two main approaches to this problem. One group of methods has addressed this by considering a fixed set of maneuvers that an actor may pursue (e.g. left, right, straight) and estimating a categorical distribution over these behaviors \cite{IntentNet, kim2017prediction, tran2015hidden}. This simplified view of the world works nicely in simulated environments, but breaks down almost immediately when driving in real cities, where intersections are complex (see Figure~\ref{fig:left-right-straight}). Another group of methods model the actor's future behavior as a multi-modal distribution over trajectories, where the set of modes is completely unstructured. These approaches have frequently suffered from the problem of mode collapse~\cite{rupprecht2017learning, multimodalDP}.

In this work, we propose a new method for predicting a distribution over the actor's future behavior that addresses the limitations of previous methods. Our approach relies on having a pre-cached map of the environment that encodes the precise topology of the road network. %\footnote{Many autonomous vehicles being developed today rely on these types of high-resolution maps.}
This topological representation captures information about the spatial relationships between lanes, thereby encoding important semantics on where and how to drive. In many existing approaches, this road topology is either ignored altogether or converted into a 2D bird's-eye view of the scene, which we posit leads to significant information loss.  Prior approaches thus under-utilize the map information and struggle with the basic task of predicting that actors will follow lanes, spurring the need for auxiliary loss functions to prevent actors from being predicted to drive off the road~\cite{ChauffeurNet}.

Our approach directly uses the mapped road topology to propose a broad set of lanes that the actor may traverse in the future. We then predict the occupancy for a sequence of spatial cells along each lane, which captures the likelihood that the actor will occupy that cell at any point over the targeted prediction horizon. This allows us to predict a distribution over the future occupancy of the actor within the mapped network of lanes. To reflect this, we call our method \emph{LaneOccupancyNet}. The key contribution of this work is a new method for encoding structure from the map into our model, which is done in such a way that (1) naturally captures the multi-modality of the future behavior of lane-following actors, (2) adapts to the local map context and generalizes to unseen road geometries, and (3) allows us to explicitly query the likelihood that a specific actor will occupy any particular lane of interest in the map, without necessarily estimating this likelihood for all lanes. These occupancy predictions can be then used determine the highest probability paths using a path-finding algorithm and generating timed trajectories or can be consumed directly by the planner to determine the cost map for the actor.
%===============================================================================
\section{Related Work}
\label{sec:related-work}

{\bf Structured Motion Prediction:} Early approaches to motion prediction focused heavily on the use of physics-based models to estimate the current motion states of actors, using a fully specified kinematic model to propagate these states into the future \cite{wan2000unscented, kalman1960new, cosgun2017towards}. Another class of structured methods leverages prior knowledge to estimate an actor's intended destinations, and then uses a planner to generate the actor's most likely trajectory to reach each goal \cite{petrich2013map, houenou2013vehicle, rehder2015goal}. % Another class of structured methods use prior knowledge to estimate an actor's intended destination (or goal), and then use a controller to plan the actor's most likely trajectory to reach its goal. to predict lane-following or maneuver-based behaviors [c].  %In addition to simulating actors using physics-based models, one can attempt to predict an actor's intended destination (goal), for instance following a specific lane, and build models of how actors perform lane-following behaviors \cite{petrich2013map, houenou2013vehicle, xie2017vehicle}. %
In contrast, our method uses a broad set of candidate goals as input to a deep network for the purpose of occupancy prediction, rather than as a prior on strict lane-following behavior. Structured methods have also been previously applied to the problem of lane occupancy prediction, for instance predicting occupancy by modeling lane-following maneuvers and actor-actor interactions using explicit policies \cite{koschi2017interaction}. Our approach also performs occupancy prediction, but with less explicit structure on motion.

{\bf Unstructured Motion Prediction:} More recently, many methods use deep learning models to directly predict future trajectories from inputs that capture the actor's state plus context from the surrounding scene \cite{luo2018fast, IntentNet, unimodalDP, multimodalDP}. In contrast to the highly structured physics-based models, these impose very little structure on the problem and instead assume the model will fully learn the patterns of actor motion from data.  While we also use a DNN to extract features from the scene, our approach differs in how the map is utilized with the model: rather than providing the map as input to an unstructured motion prediction system, it is used to directly query occupancy probabilities for relevant spatial locations.

{\bf Intention Prediction:} In addition to directly predicting motion, inferring an actor's higher-level intent can help predict its future behavior. Intent can be modeled as a discrete set of actions, and several methods directly predict these action categories \cite{kumar2013learning, streubel2014prediction, IntentNet}or by producing multimodal trajectory predictions with each mode representing a particular semantic category of behavior \cite{multimodalDP, makansi2019overcoming}. Our work focuses specifically on spatial multimodality, and we use structure from nearby mapped lanes to enumerate maneuvers and predict future occupancy without restricting ourselves to a pre-determined set of action categories.

\section{APPROACH}
\begin{figure}
\centering
\begin{minipage}[b]{\linewidth}
\centering
\includegraphics[width=\linewidth]{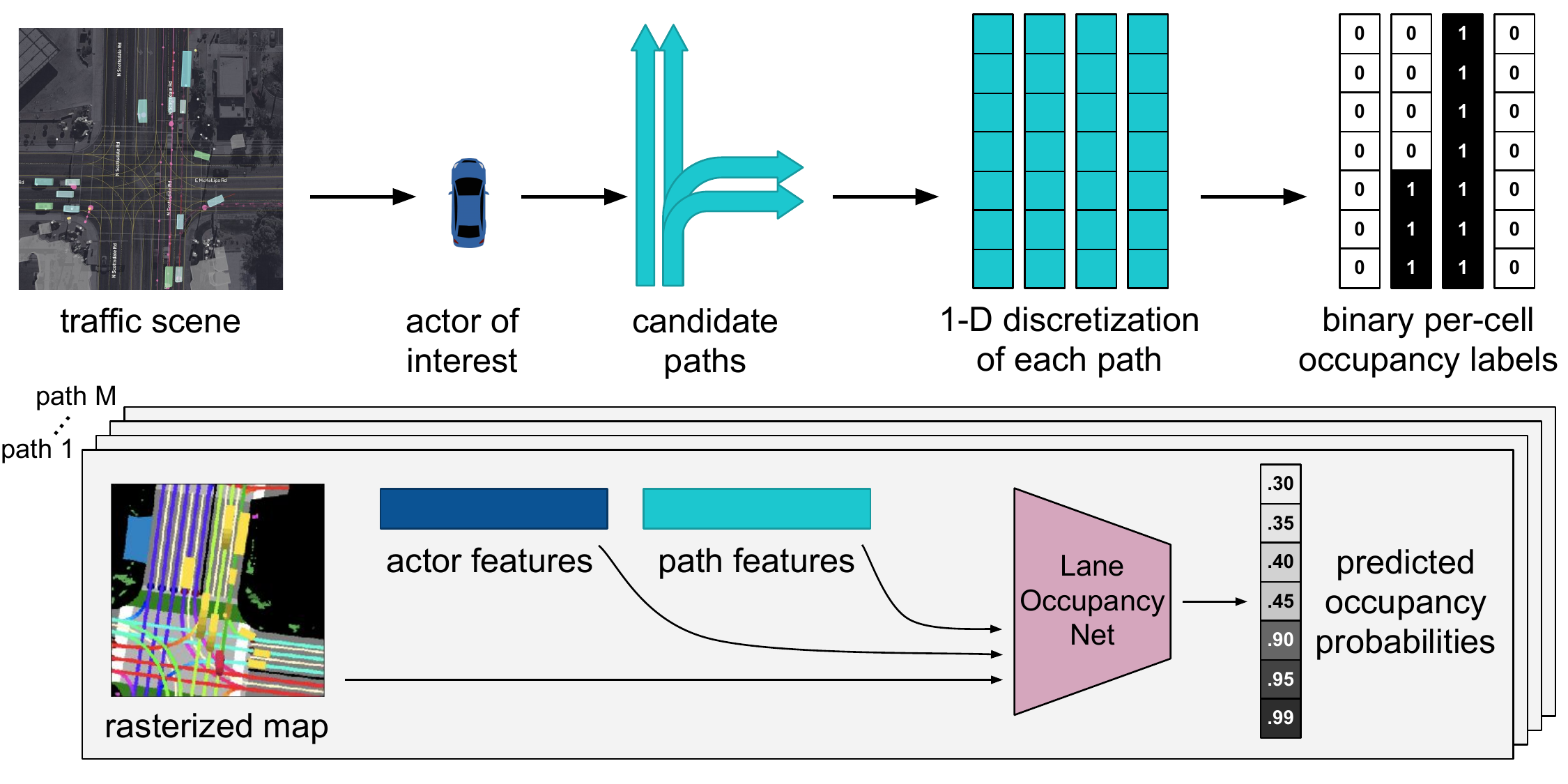}
\end{minipage}
\caption{An overview of our approach. Top: For a given traffic scene, we identify an actor of interest, generate a set of candidate lane paths for the actor, discretize the path into cells, and label each cell according to the actor's ground truth occupancy. Bottom: We process each path separately and apply our LaneOccupancyNet to predict the spatial occupancy probabilities along the path.}
\vspace{-.2cm}
\label{fig:approach-overview}
\end{figure}

In this section, we describe our approach for predicting vehicle occupancy over the lane network. In order to focus on the prediction problem, we assume we already have an object detection and tracking system which operates on sensor data to produce a list of actors in the scene along with their state estimates. We further assume that we have access to a high-resolution map of our surroundings, which encodes road boundaries, lane boundaries, lane connectivity, lane directions, crosswalks, traffic signals, and other details of the scene geometry. In this approach, we consider each actor in turn and generate predictions for the actor of interest in the context of all other actors in the scene, including the self-driving vehicle (SDV) itself. An overview of our approach is shown in Figure~\ref{fig:approach-overview}.

\subsection{Path Generation}
The first step in our approach is to use the map to generate a set of candidate paths that the actor may follow.

\begin{figure}
\centering
%\begin{minipage}[t][.1cm]{0.03\linewidth} 
%(a)
%\end{minipage}
\begin{minipage}[t]{0.3\linewidth} 
\includegraphics[width=\linewidth, trim={3mm 3mm 3mm 3mm, clip}]{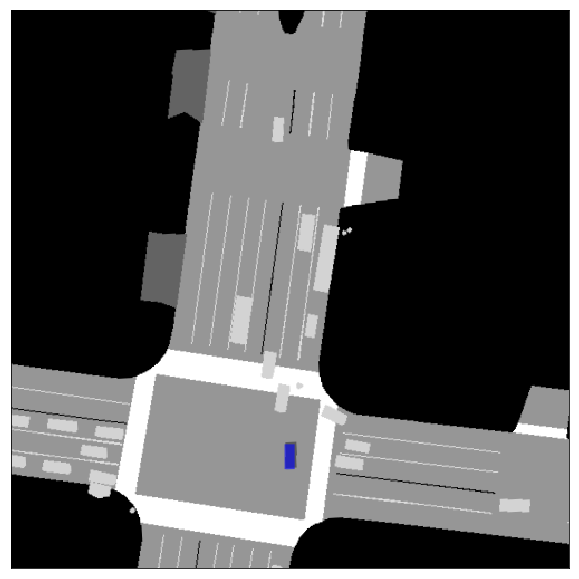}
\end{minipage}
\hspace{.1cm}
%\begin{minipage}[t][.1cm]{0.03\linewidth} 
%(b)
%\end{minipage}
\begin{minipage}[t]{0.3\linewidth}
\includegraphics[width=\linewidth, trim={3mm 3mm 3mm 3mm, clip}]{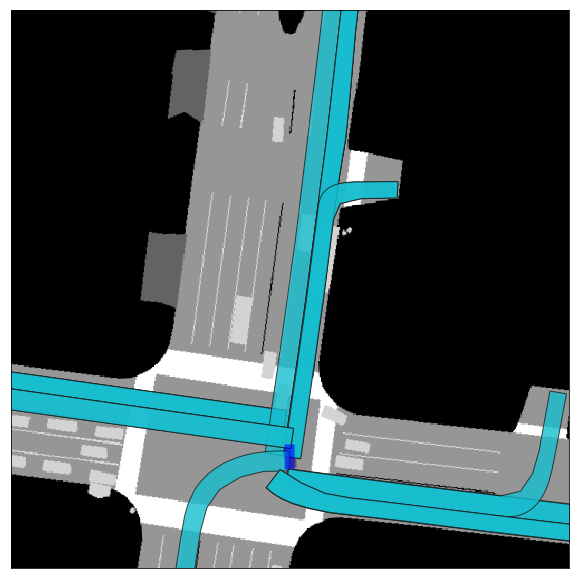}
\end{minipage}
\hspace{.1cm}
%\begin{minipage}[t][.1cm]{0.03\linewidth} 
%(c)
%\end{minipage}
\begin{minipage}[t]{0.3\linewidth}
\includegraphics[width=\linewidth, trim={3mm 3mm 3mm 3mm, clip}]{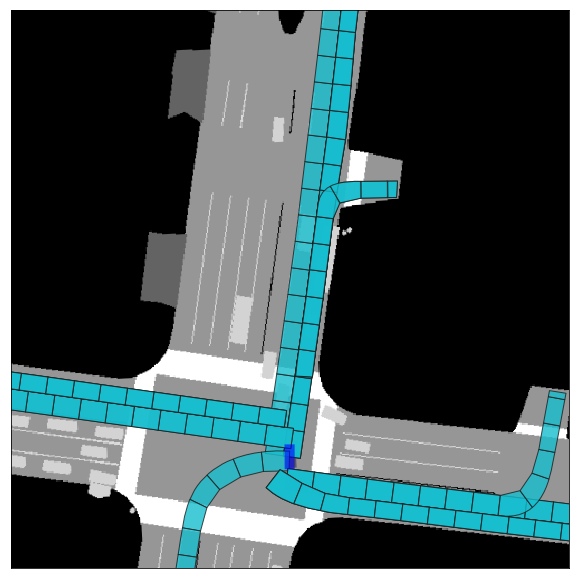}
\end{minipage}
\caption{Left: A scene with the actor of interest shown in dark blue. Middle: The set of candidate lane paths generated for the actor of interest. Right: The discretization of each path into fixed-length cells.}
\vspace{-.2cm}
\label{fig:path-proposal}
\end{figure}

We define each path as a region in 2D space that is specified by a sequence of lanes with no branching. Lanes that branch will be split into multiple paths that diverge at the branching point. To generate a set of paths for the actor of interest at a given time $t$, we query the map to identify all lanes that fall within 2 meters of the actor's position $\psi_t$. Then, starting from the point $\psi_t$, we ``roll out'' the paths according to the lane successor relationships specified by the map, up to a fixed distance. This yields a collection of candidate paths for the actor as shown in the middle image in Figure~\ref{fig:path-proposal} for an actor in the intersection.

Together, the spatial area covered by the union of the paths determines the region over which we predict the occupancy of the actor. One potential drawback of this approach is that we explicitly do not predict other actor's occupancy over regions of the world that are not covered by mapped lanes. However, it's important to note that areas outside of mapped lanes are specifically of \emph{less} interest to us, as we assume the SDV is designed to follow typical rules of the road and drive only within mapped lanes. Thus, predicting the occupancy of  actors within these regions is of much higher importance.

Once we have the set of candidate paths, we discretize each path into a fixed length sequence of cells, where the number of cells $L$ is determined by the cell length. This discretization enables us to capture maneuvers like lane changes where the actor does not stay in the same lane for the duration of the prediction horizon. We constrain the discretization to be one-dimensional, meaning that each cell will cover the full width of the lane. We make this choice based on an observation that the SDV cares much more about which lanes will be occupied than it cares about the precise lateral position of a vehicle within the lane. Finally, given these cells, we label each one according to whether or not the vehicle's polygon entered that cell at any point over the prediction horizon $H$.

\subsection{Occupancy Prediction}
\label{sec:occupancy-prediction}

Given the set of candidate paths and the sequence of cells along each path, we aim to predict whether or not each of these cells will be occupied by the actor. Because we are predicting spatial occupancy rather than spatio-temporal occupancy, the actor can occupy multiple cells over the duration of the prediction horizon. As a result, we want to predict an occupancy probability in the $[0, 1]$ range for each cell rather than a normalized categorical distribution over all cells. To do this, rather than jointly predicting the occupancy for all cells at once, we consider each path independently and predict the occupancy only over the cells in a single path. 

Specifically, let %$O_{k} \sim \textrm{Bernoulli}(p_{k})$ 
$O_1, \cdots, O_L$ be a sequence of binary random variables where $O_k$ indicates whether the $k$-th cell in a path was occupied at any point over the next $H$ seconds. We assume that our data consists of independent samples from the joint distribution $p(O_1, \cdots, O_L)$ parameterized by the per-cell occupancy probabilities $p_1, \cdots, p_L$. Our goal then is to estimate
\begin{equation*}
\hat{p}_1, \cdots, \hat{p}_L = \hat{f}(X)
\end{equation*}
where $X$ are the inputs to our model. This approach, in which we process each path separately, has several benefits. First, it allows us to consider an arbitrary number of paths for each actor without relying on truncation or padding to force the paths into a fixed-size output representation. This flexibility in the number of paths enables our method to adapt to the local map context. As an example, an actor driving on a single-lane road far from any intersection will only have a single candidate path. In contrast, an actor approaching the 6-way intersection depicted in Figure~\ref{fig:left-right-straight} may have 10 or more candidate paths. Another benefit of this approach is that it provides a path-centric output representation. This enables our model to generalize  very well to unseen road geometries, as long as the map accurately captures the lane topology.

% To capture this, we do not model the problem as a categorical distribution over cells but instead as a joint distribution over a sequence of binary random variables that indicate the spatial occupancy of each cell in the sequence.

%Let $O_{j} \sim Bernoulli(p_{j})$ be a random variable that indicates whether the $j$-th cell in a path was occupied at any point over the next $H$ seconds, where $H$ denotes the prediction horizon. We assume that our data consists of independent draws from the joint distribution of $O_1, \cdots, O_L$. 

\subsection{Labels}
\label{sec:labels}

For a given actor $i$ and a given path $j$, we assign a binary label $Y_{ijk} \in \{0, 1\}$ to each cell $k$ along the path. To determine the label, we use the future ground truth trajectory of the vehicle truncated at the prediction horizon. If the vehicle's ground truth polygon touches the cell at any point over the horizon, we label it 1, and otherwise we label it 0. 
%For a given prediction horizon $H$, if any point in the vehicle's ground truth polygon touches a cell, we give the cell a \textit{True} label else we label it \textit{False}. We define occupancy as a non-zero overlap between the vehicle's bounding box and the discretized cell polygon. For each cell we find the point in time in the prediction horizon where the vehicle is closest to the center of the cell and use the vehicle's bounding box estimate at this instance for determining the label for the cell. 
When we observe the vehicle for a duration ${h}$ shorter than the prediction horizon ${H}$, we only know the positive labels for certain (the other cells may or may not be visited in the remaining $H-h$ seconds). For these cases, we label all cells the ground truth polygon does not touch with a sentinel value of -1. These cells are ignored in the loss function, and therefore are not used for either training or evaluation. Since positive labels are scarce relative to negative labels, this allows us to leverage additional positive samples for training.

\subsection{Model}
\label{sec:model}

In this section we provide an overview of our occupancy prediction model, LaneOccupancyNet (LON), including a description of the input representation, the network architecture that is inspired by \cite{2dFusionDP}, and the output representation. The overall model architecture is shown in Figure~\ref{fig:model-architecture}.

\subsubsection{Input Representation}
\label{sec:input-representation}

There are three different inputs we provide to our model to capture different pieces of information that influence an actor's future occupancy: context from the actor's neighborhood, information about the actor's current and past behavior, and information about the candidate goal path.

\emph{(a) Scene rasters:} The actor's future behavior is influenced by the geometry of the scene and the positions of other nearby vehicles. To capture this context, we provide a rasterized RGB image capturing a bird's-eye view representation of the scene at the current time instance, oriented based on the position and heading of the actor of interest. The rasters have a resolution of 0.2m and capture a 60m x 60m region, with 10m behind the actor and 50m in front of it. These rasters are similar to the ones used in \cite{multimodalDP} with the addition of the candidate path overlayed on the raster in dark green, to capture the path along which we are predicting occupancy. 

\emph{(b) Actor features:} 1D array of hand-crafted features that capture the actor's current state and past behavior, such as the actor's speed, its angular velocity, and the variance in its heading for the past 3 seconds. %Since the actor state estimates are computed from the SDV's sensor observations, and the accuracy of these estimates depend to some degree on the actor's relation to the SDV, we also include features that capture the actor's relationship to the SDV such as how far the actor is from the SDV, their speed with respect to the SDV, etc. %Hence in addition to features that are derived just from the actor's state estimates, we also provide hand-crafted features that capture the relationship of the actor with the SDV.

\emph{(c) Path features:} 1D array of hand-crafted features that capture additional information about the candidate path. Since the scene raster shows only an early portion of the candidate path, these features help provide information about the entire path, such as its curvature. We also provide information about the actor's relationship to the path, such as the path-relative position, velocity, acceleration, and heading, along with the history of these values cached from previous cycles.

\subsubsection{Network Architecture}

\begin{figure}
\centering
\begin{minipage}[b]{\linewidth}
\centering
\includegraphics[width=\linewidth]{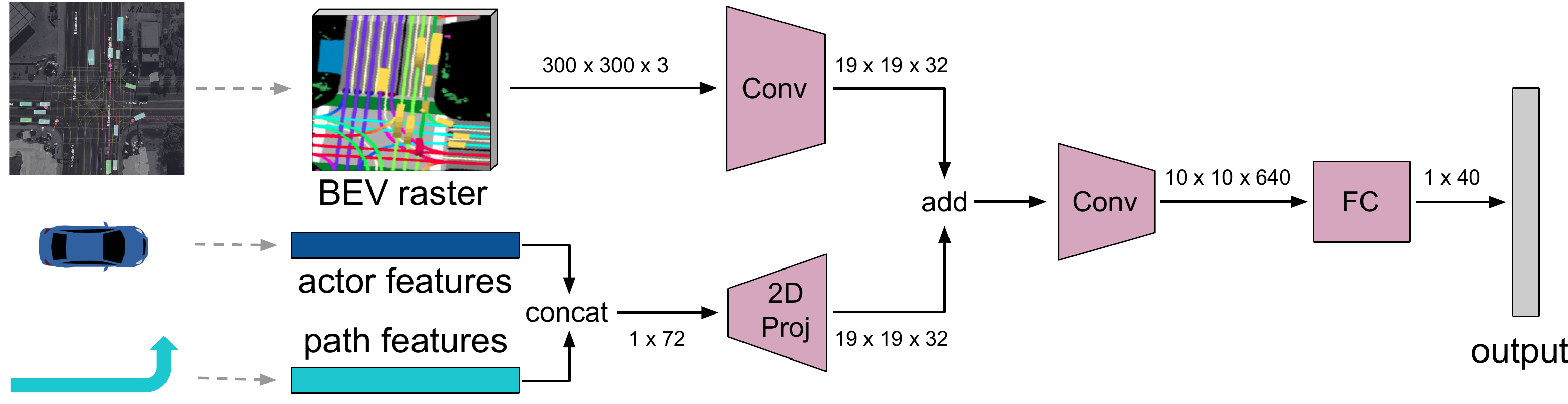}
\end{minipage}

\caption{The architecture of LaneOccupancyNet, which predicts 1-D occupancy along a path. We extract scene features from the BEV raster by passing the image through a convolutional network. We then fuse these with the 1D actor and path features by projecting them into the 2D space of the scene features using ideas from~\cite{2dFusionDP}. The projection operation uses a sequence of a fully connected layer, a reshape, and a 1x1 convolutional layer. Lastly, we have a second convolutional block and a fully connected block with 2 hidden layers of sizes 2048 and 1024, followed by the output layer.}
\vspace{-.2cm}
\label{fig:model-architecture}
\end{figure}

Our model architecture is inspired by the FastMobileNet architecture proposed in~\cite{2dFusionDP}. In particular, we project the 1D concatenated array of actor and path features into the 2D space of the latent scene features. This allows us to directly add the projected features to the scene features and perform additional convolutional operations on top. As pointed out in \cite{2dFusionDP}, we hypothesize that this form of feature fusion allows the map information at different spatial locations to interact differently with the engineered features. See Figure~\ref{fig:model-architecture} for details. %The network is made up of convolutional layers for extracting image features followed by fully connected layers that combine the features from all three input streams into a one-dimensional output vector. For efficiency, we use the MobileNet V2 architecture for our Base CNN \cite{MobileNet}. The first two fully connected layers have 2048 and 1024 output nodes, respectively. These are followed by another fully connected layer that produces the output vector containing one value per cell.

\subsubsection{Output Representation}

We independently predict the future spatial occupancy of each actor along each of its paths, with the $k$-th element in the output representing the probability of the actor occupying the $k$-th cell along the path for any duration within the entire prediction horizon. We use the sigmoid cross entropy loss for each cell and compute the mean loss over all cells. %Letting $Y_{ijk}$ denote the ground truth binary label for a given cell and $\hat{Y}_{ijk}$ denote the predicted occupancy, the complete loss function is given by:
%\begin{equation*}
%L(Y, \hat{Y}) = \sum_{\text{actor}\,i} \sum_{\text{path}\,j} \sum_{\text{cell}\,k} ~ \mathbb{I}\{Y_{ijk} > -1\} ~\ell(Y_{ijk},\hat{Y}_{ijk}) 
%\end{equation*}
%where $\ell$ is the standard log loss.

%{\small
%\begin{align*}
%L(Y, \hat{Y}) = & \sum_{\text{actor}~i} \sum_{\text{path}~j} %\sum_{\text{cell}~k} ~ \mathbb{I}\{Y_{ijk} > -1\} \\
%               & \left[ Y_{ijk}  \log(\hat{Y}_{ijk}) + (1-Y_{ijk})  \log(1-\hat{Y}_{ijk}) \right]
%\end{align*}
%}%

\subsection{Implementation Details}
\label{sec:implementation}

We train our model for 50,000 iterations as we observe the validation loss stabilizes by then. We set the learning rate to \(10^{-4}\) with a decay of 0.9 every 11000 steps. It takes around 12 hours to train the model using distributed training on 4 GPUs. In practice, we use a cell length of 4.8 meters, since this is the length of the average car, and use 40 cells per path, for a path length of 192 meters (this allows us to capture fast-moving actors). We experiment with prediction horizons of 3, 6, and 9 seconds, but primarily report results with the 9-second horizon.

%Note that that the set of paths we consider will likely have some spatial overlap (e.g., if a lane branches into 3 successor lanes, we will create 3 separate paths that partially overlap). To handle this, we also perform a post-processing step to estimate the final occupancy of each unique cell by taking the average of the set of estimates from each path that contains that cell.
\section{EXPERIMENT RESULTS}
\subsection{Dataset}

For training and evaluation, we use the large-scale ATG4D dataset described in \cite{meyer2019lasernet}, which contains a variety of interesting scenarios and diverse driving behaviors from multiple cities across North America. The training and validation sets contain 5,000 and $500$ 30-second scenarios, respectively. To train our model, we randomly sample 5\% of the training set. %A single training sample comprises one candidate path for a given vehicle at a given instance of time.
We report results on the validation set, which contains 127,669 actor frames with at least 9s of observed future. For all experiments, we measure performance only on moving vehicles (those with estimated speed $>$ 0.5 m/s) that are within 50m of the SDV and are observed for at least the full duration of the prediction horizon in the future.
%Our validation set contains a total of ~311,500 samples, of which 127,669 have at least 9 seconds of observed future. 

\subsection{Spatial Occupancy Metrics}
\label{sec:spatial-occupancy-metrics}

We evaluate our system performance by comparing an actor's true spatial location against the spatial occupancies predicted by different methods using an average likelihood metric.  We compare against two baselines: (1) unimodal trajectories generated by an unscented Kalman filter (UKF) \cite{wan2000unscented} that forward-propagates actor states from a second order tracking system; (2) trimodal trajectories generated by an unstructured deep network \cite{multimodalDP}, identified here as Multiple Trajectory Prediction (MTP). The MTP model was trained on the same data as our proposed method. In both baseline methods, each predicted trajectory consists of a sequence of waypoints over time, position covariances surrounding each waypoint, and probabilities per trajectory (relevant in the multimodal case). In contrast, our method predicts spatial occupancy up to a future time horizon. 

\begin{figure}[b]
\centering
\begin{minipage}[t]{0.24\linewidth}
\centering
{\small Ground Truth}\\[2pt]
\includegraphics[width=\linewidth]{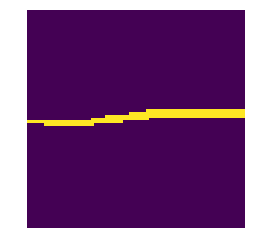}%\\[-7pt]
%{\small Ground Truth}
\vspace{-.1cm}
\end{minipage}
\hfill
\begin{minipage}[t]{0.24\linewidth}
\centering
{\small UKF}\\[2pt]
\includegraphics[width=\linewidth]{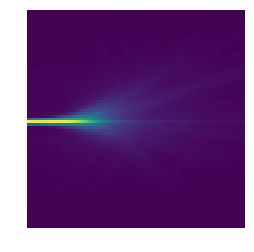}%\\[-7pt]
%{\small UKF}
\vspace{-.1cm}
\end{minipage}
\hfill
\begin{minipage}[t]{0.24\linewidth}
\centering
{\small MTP}\\[2pt]
\includegraphics[width=\linewidth]{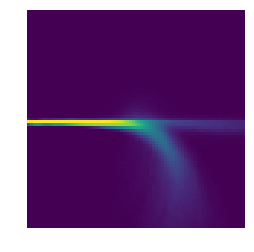}%\\[-7pt]
%{\small MTP}
\vspace{-.1cm}
\end{minipage}
\hfill
\begin{minipage}[t]{0.24\linewidth}
\centering
{\small LON}\\[2pt]
\includegraphics[width=\linewidth]{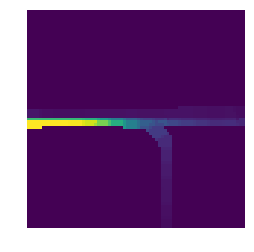}%\\[-7pt]
%{\small LON}
\vspace{-.1cm}
\end{minipage}
\caption{
Ground truth and predicted occupancy heatmaps, shown for each method. In each figure, the actor starts on the left and then moves to the right. The UKF predicts a unimodal trajectory that disperses at future horizons. MTP predicts a trimodal distribution of trajectories that describe multiple possible future motions. Our method, LON, produces likelihoods tied to the geometry of the nearby lanes.}
\label{fig:occupancy}
\end{figure}

In order to compare these two different representations, we first convert the output of all methods into a common representation that consists of 2D spatial occupancy predictions over a grid that is centered at the actor's current position (we use a 150m x 150m grid with 1m resolution). Ground truth labels are generated by determining which 2D cells an actor occupied at any point over the prediction horizon. Figure~\ref{fig:occupancy} shows an example of the ground truth occupancy mask and predicted occupancy likelihoods. %We first generate ground truth labels by identifying all 2D locations that an actor occupies over the prediction horizon. Next, for UKF and MTP, we use Monte Carlo sampling to estimate occupancy likelihoods. We draw samples from the possible trajectory modes (unimodal for UKF, trimodal for MTP) and from the waypoint covariances to estimate the likelihood for each 2D location.

Given a ground truth 2D occupancy grid and a corresponding predicted likelihood grid, we compute the overall average likelihood as 
\begin{align*}
    P(Z \vert \hat{Z}) &= \frac{1}{G^2} \sum_{ij} P(Z_{ij} \vert \hat{Z}_{ij} ) \\
    &= \frac{1}{G^2} \sum_{ij} (\hat{p}_{ij}^{Z_{ij}} + (1 - \hat{p}_{ij})^{1 - Z_{ij}}).
\end{align*} 
Note that $Z_{ij}$ is the ground truth label of the $i,j$-th cell in 2D space (as opposed to in path space), $\hat{p}_{ij}$ is the predicted occupancy likelihood for that cell, and $G$ is the grid size. We also compute two additional metrics, the positive likelihood, which is calculated only on 2D cells with label $Z_{ij} = 1$, and the negative likelihood, which is calculated only on 2D cells with label $Z_{ij} = 0$. 

Next we describe how to convert the two output representations into the common 2D spatial representation.

\subsubsection{2D Spatial Occupancy from Trajectories} 
A multimodal trajectory prediction is defined as a time-varying mixture of Gaussian distributions. Each trajectory in the mixture of $K$ components is given by $f_k(t) = \mathcal{N}(\mu_k(t), \Sigma_k(t))$,
%    f_k(t) = \mathcal{N}(\mu_k(t), \Sigma_k(t))
%\end{equation}
where $\mu_k(t) \in \mathbb{R}^2$ is the mean 2D position of the actor at time $t$ and $\Sigma_k(t)$ is the corresponding position covariance. Each trajectory also has an associated mode probability, $p_k$. In a unimodal case, such as UKF, $K = 1$ and $p_1 = 1.0$.

To convert a spatio-temporal trajectory into a spatial occupancy likelihood, for each 2D location, we determine the probability that an actor ends up occupying this location at any time over the prediction horizon. To approximate this, we utilize a Monte Carlo sampling technique. We first generate $N$ samples by repeating the following procedure:
\begin{itemize}
    \item Sample a mode from the distribution over the mixture components $k \sim \text{Categorical}(p_1, \cdots, p_K)$. % set of trajectory predictions, based upon the normalized distribution of discrete probabilities, $p_k$.
    \item Sample a trajectory as follows. First, sample from a 2D standard Gaussian distribution $z \sim \mathcal{N}(0, I)$. Then, for each time point $t$, convert $z$ into a sample from $\mathcal{N}(\mu_k(t), \Sigma_k(t))$. To do this, we compute $A_k(t)$, the Cholesky decomposition of $\Sigma_k(t)$, and then calculate $x_t = A_k(t) \, z + \mu_k(t)$. Repeat this for all time points to obtain a sequence of sampled positions $x_n = x_1, \cdots, x_T$. This allows us to sample a coherent trajectory from the sequence of Gaussians.
    %\item For a given mode, sample $f_n(t)$ from $\mathcal{N}(\mu_k(t), \Sigma_k(t))$.\footnote{In lieu of applying a full vehicle model to estimate all possible actor motions within the multivariate distribution, the set of sampled  actor trajectories can be thought of as a fiber bundle surrounding the mean trajectory.  While this approximation does not accurately capture possible actor motions, it is sufficient for purposes of our occupancy likelihood sampling.}
    \item Define $P_n$ as the swept volume produced by the actor's polygon moving along the sampled trajectory $x_n$.%$f_n$.
\end{itemize}
Finally, to estimate the predicted occupancy likelihood at cell $i,j$ in the 2D grid, we simply check each cell against the swept volumes from each of the sampled trajectories, and calculate the frequency with which the cell is occupied:
\begin{equation*}
    \hat{p}_{ij} = \frac{1}{N} \sum_{n=1}^N
      \begin{cases}
      1, & \text{if } P_n \text{ overlaps with cell i,j} \\
      0, & \text{otherwise}
      \end{cases}
\end{equation*}
In our experiments, we use $N=1000$.%, and we represent the world as a discretized 2D grid of resolution 1.0 meters with equal width and height of 150 meters, centered on the actor of interest.

\subsubsection{2D Spatial Occupancy from Path-Based Occupancy} Since LaneOccupancyNet directly predicts the spatial occupancy probability for cells along each path, we directly use the polygon shapes of the cells to map these into the 2D likelihood grid. Note that that the set of paths we consider will likely have some spatial overlap (e.g., if a lane branches into 3 successor lanes, we will create 3 separate paths that partially overlap). To handle this, we also perform a post-processing step to estimate the final occupancy of each unique cell by taking the average of the set of estimates from each path that contains that cell.

\subsection{Multimodality Estimation}

Since actors choose over multiple future actions and we are interested in prediction methods that capture these different possibilities, we also developed a measure to evaluate the spatial multimodality of each prediction method.

Our measure calculates the unique number of spatial modes an actor may follow according to its predicted occupancy. Modes are defined as spatially distinct paths to get from an actor's current location out to a new location some distance away. To count the modes, we trace 1D likelihoods along concentric rings at varying ranges from the actor's current location, and count the number of observed peaks in the likelihood along these rings.

\begin{figure}[t]
    \centering
    \begin{minipage}[b]{0.4\linewidth}
    \centering
\includegraphics[width=0.48\linewidth]{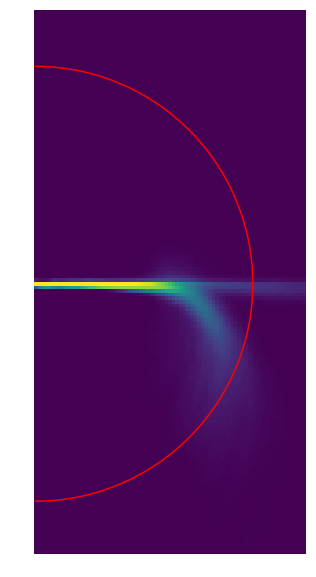}
\includegraphics[width=0.48\linewidth]{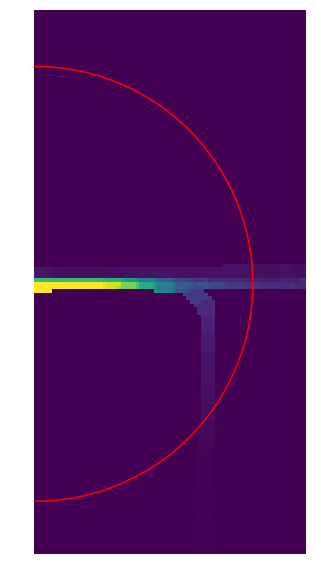} %\\ (a)
    \end{minipage}
    %\hspace{.025cm}
    \hfill
    \begin{minipage}[b]{0.58\linewidth}
    \centering
\includegraphics[width=\linewidth]{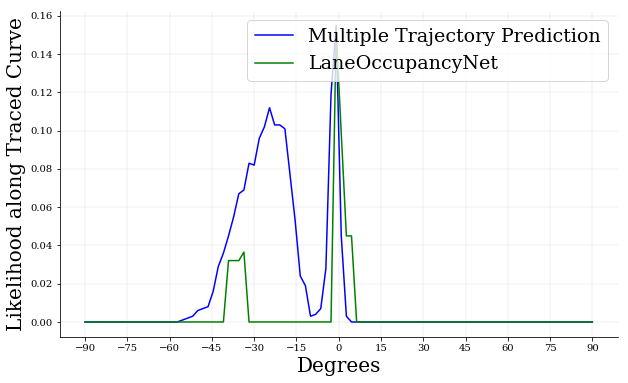} %\\ (b)
    \end{minipage}
\caption{A visual overview of the multimodality measure. Left: Spatial likelihoods for the MTP and LON approaches. In each, the spatial occupancies clearly show two separate possible paths. The traced curves show a 180 degree arc in the forward facing direction of the actor of interest. Right: Plot showing 1D likelihoods for the arcs traced on the left. In each case, we see two obvious peaks. Our multimodality estimation method uses peak finding to count the number of spatial modes at each range.}
\label{fig:multimodality_estimation}
\vspace{-.2cm}
\end{figure}

Figure~\ref{fig:multimodality_estimation} provides a visual description of this general approach. Once 1D likelihoods are obtained for a ring at a given distance from the actor, the estimated number of modes is determined by counting the number of peaks in each curve. A peak is defined as a local maximum that rises at least $\tau$ above neighboring local minima.  In our experiments, we use $\tau = 0.03$, and focus on the 180 degree arcs in the forward direction of the actor of interest, using its estimated heading.

\subsection{Quantitative Results}

\begin{table}[b]
\caption{Average likelihood results for the two baselines vs.~our proposed method (9s horizon).}
\centering
\begin{tabular}{l|c|c|c}
\toprule
Method & Overall & Positive & Negative \\
\midrule
Unscented Kalman Filter & 0.9873 & 0.3709 & 0.9911 \\
Multiple Trajectory Prediction & \textbf{0.9955} & 0.6577 & \textbf{0.9976}\\
LaneOccupancyNet & 0.9934 & \textbf{0.7482} & 0.9949 \\ 
\bottomrule
\end{tabular}
\label{tab:likelihoods}
%\vspace{-.6cm}
\end{table}

%%% OLD (b):
% The multimodality of each method (9s horizon) is estimated at different ranges from the actor's location. Both the UKF and MTP consistently exhibit mostly unimodal predictions, despite the latter being a trimodal model. LaneOccupancyNet consistently proposes multiple possible paths for each actor.

%%Right: Comparison of average positive likelihood for models with different prediction horizons. As we increase the horizon, our model does significantly better than the baselines.

\begin{figure}[tbp]
    \centering
%    \begin{minipage}[b]{0.487\linewidth}
    \begin{minipage}[b]{0.55\linewidth}
    \centering
    \includegraphics[width=\linewidth, trim={2cm 0cm 2cm 0cm, clip}]{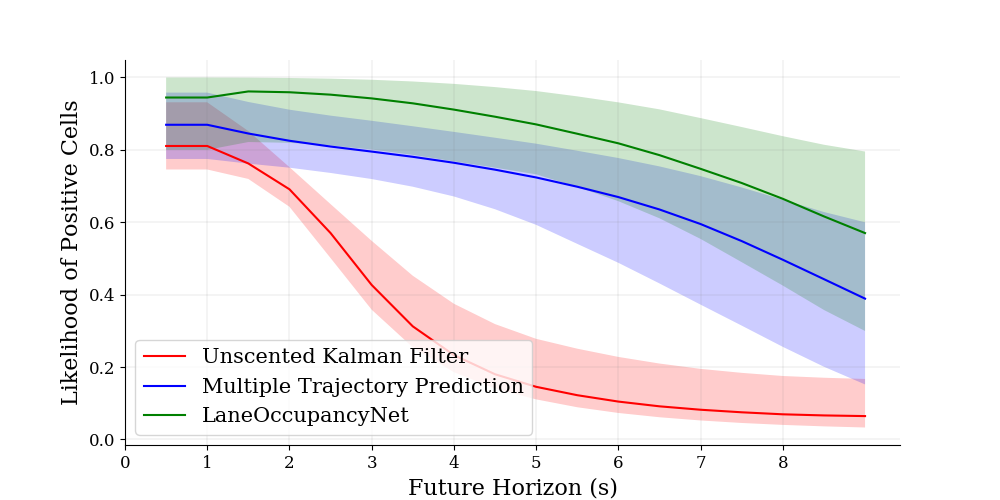} \\ (a)
    %\label{fig:spatial_occupancy_per_horizon}
    \end{minipage}
    \hfill
%    \begin{minipage}[b]{0.48\linewidth}
    \begin{minipage}[b]{0.43\linewidth}
    \centering
    \includegraphics[width=\linewidth, trim={2cm 0cm 2cm 0cm, clip}]{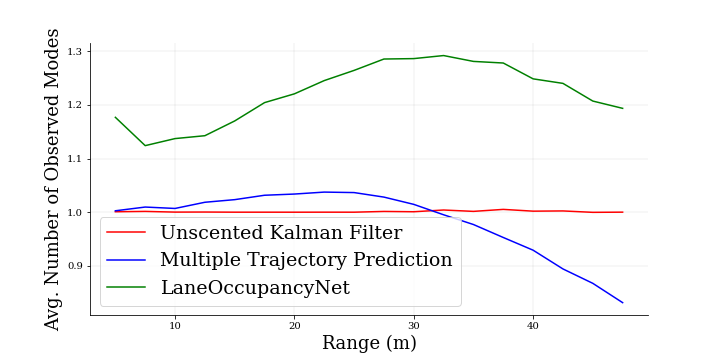} \\ (b)
    %\label{fig:multimodality}
    \end{minipage}
\caption{(a) Comparison of the positive likelihood distribution across the 3 methods shows that LON best predicts occupied areas at future time horizons.  %Median likelihoods, along with 25th and 75th percentiles are shown.%Comparison of the distribution of positive likelihood across 3 methods. Our approach more consistently predicts occupied areas many seconds into the future. 
Solid lines show the median average likelihood across the scenes %(one per actor per time frame)
in our test set, shaded regions corresponding to 25th and 75th percentiles.
(b) Observed number of spatial modes at varying distances.  UKF is unimodal by design.  MTP exhibits multimodality at shorter range, diminishing at longer ranges due to short predictions in some cases. LON has a greater number of modes overall. %Estimated multimodality of each method (on a 9s horizon) across different ranges. The UKF is unimodal by design, as is observed here.  MTP exhibits multimodality at shorter range, however further ranges tend to not have as many detected modes.  LaneOccupancyNet has a greater number of modes overall and increases multimodality at range, via multiple distinct paths per actor.}
}
\vspace{-.2cm}
\label{fig:occupancy_and_modality_plot}
\end{figure}

Table~\ref{tab:likelihoods} shows comparative results on the three likelihood measures. While the overall likelihood and negative likelihood both degrade with our method, we observe a 13.8\% improvement in positive likelihood. This suggests that our approach does a better job of estimating all of the possible places an actor may end up, effectively a measure of recall for prediction methods. We explore this further by plotting the positive likelihood at different future time horizons of the ground truth. We measure actor occupancy across individual points of time and compare against our predictions. As seen in Figure~\ref{fig:occupancy_and_modality_plot}(a), our method consistently outperforms both UKF and MTP on this metric.  %Additionally, by showing percentiles around these estimates, we see a large number of occupancy predictions extremely close to the perfect 1.0 using our method across horizons, as compared to the baseline which steadily decreases with larger and larger time horizons.

%We also report results on the multimodality measure that we described earlier.
%Since actors choose over multiple future actions, we developed a measure to evaluate the spatial multimodality of each prediction method as described earlier. %To do this, we count the number of local maxima along 1-dimensional concentric rings measured over each 2D likelihood grid.
Using the approach described earlier, we evaluate the multimodality of all three methods over a set of ranges, and plot the results in Figure~\ref{fig:occupancy_and_modality_plot}(b). These results demonstrate that observe a greater degree of multimodality in the predictions from LaneOccupancyNet than those from UKF and MTP. %, supporting the case that our method is better able to adapt to complex scenes. %As an advantage of our approach is ability to adapt to more complex scenes, this analysis demonstrates a greater degree of multimodality.
Given that a key advantage of our approach is that we are able to adaptively add modes to our predicted occupancy distribution as additional lanes appear in the scene, this analysis supports the argument that we are better able to handle complex scenes where actors can choose over many different paths.

\subsection{Qualitative Results}

\begin{figure}[!ht]
\centering
\begin{minipage}[c]{0.43\linewidth}
{\begin{flushleft} \small ~ Truth ~~~ MTP ~~~  LON  \end{flushleft}}
\vspace{-1mm}
\centering
{\includegraphics[width=\linewidth, trim=4.7cm 1.6cm 3.7cm 1.8cm, clip]{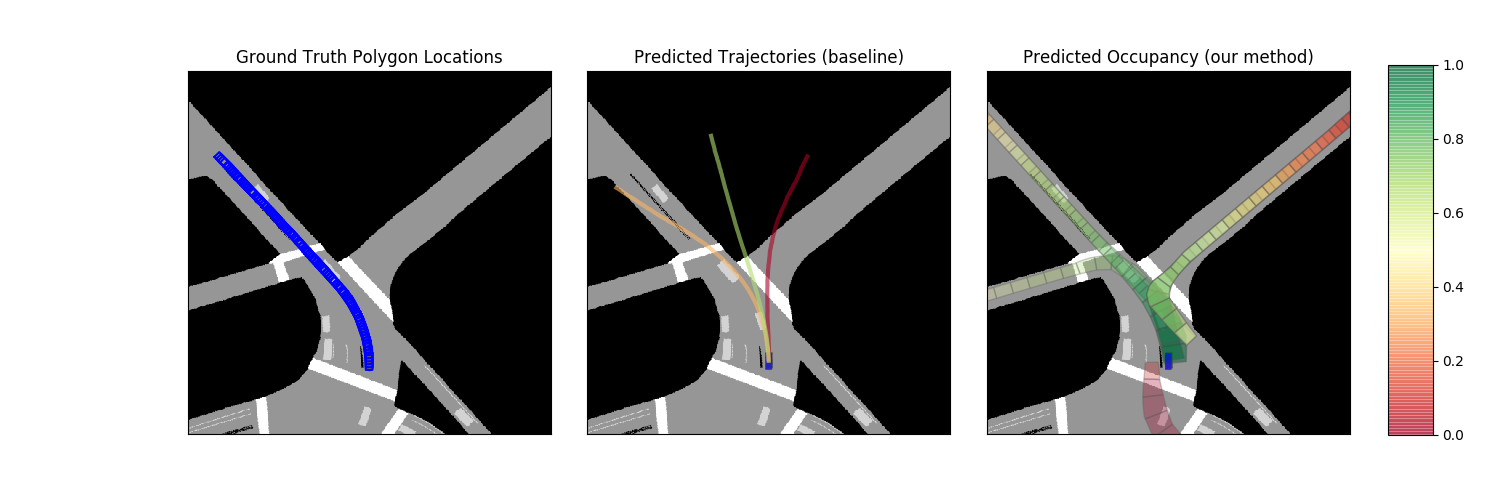}}\\[3pt]
{\includegraphics[width=\linewidth, trim=4.7cm 1.6cm 3.7cm 1.8cm, clip]{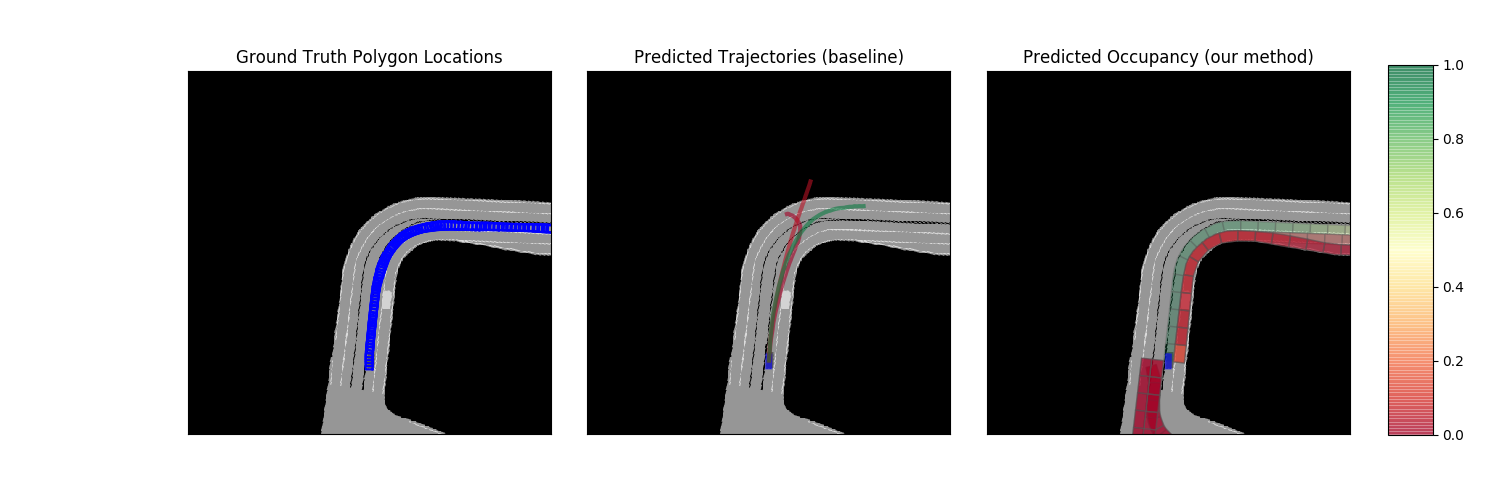}}\\[3pt]
{\includegraphics[width=\linewidth, trim=4.7cm 1.6cm 3.7cm 1.8cm, clip]{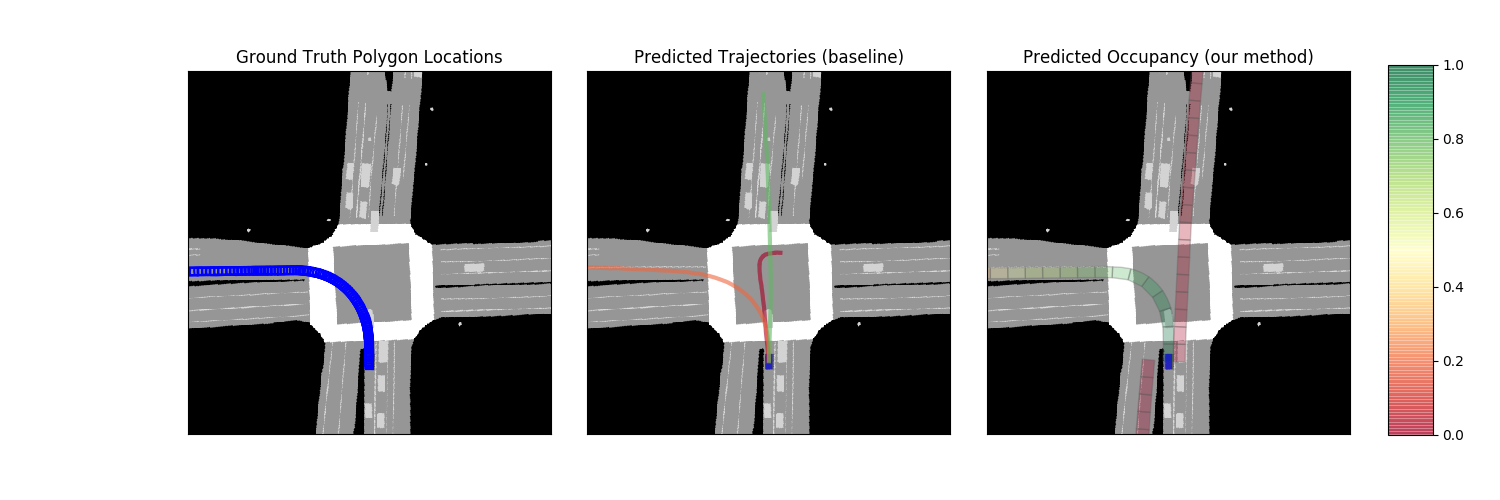}}\\[3pt]
{\includegraphics[width=\linewidth, trim=4.7cm 1.6cm 3.7cm 1.8cm, clip]{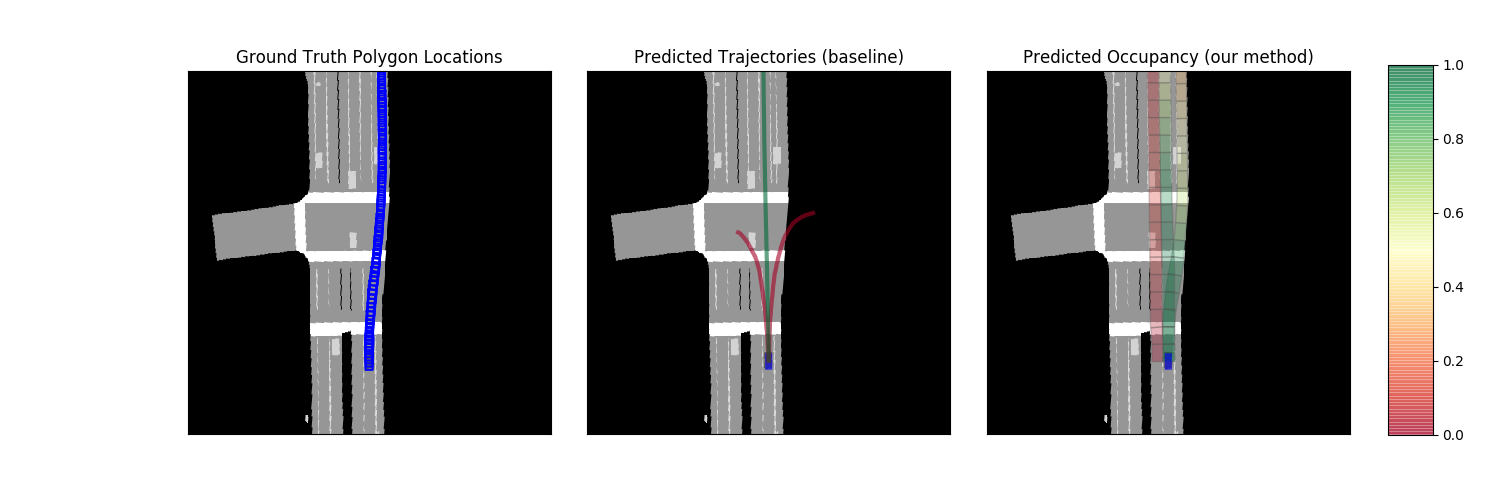}}\\[3pt]
{\small (a) Top Cases}
\end{minipage}
%\hfill
\begin{minipage}[c]{0.06\linewidth}
\centering
\includegraphics[width=\linewidth]{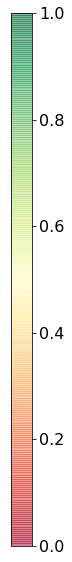}
\end{minipage}
%\hfill
\begin{minipage}[c]{0.43\linewidth}
{\begin{flushleft} \small ~ Truth ~~~ MTP ~~~  LON \end{flushleft}}
\vspace{-1mm}
\centering
\includegraphics[width=\linewidth, trim=4.7cm 1.6cm 3.7cm 1.8cm, clip]{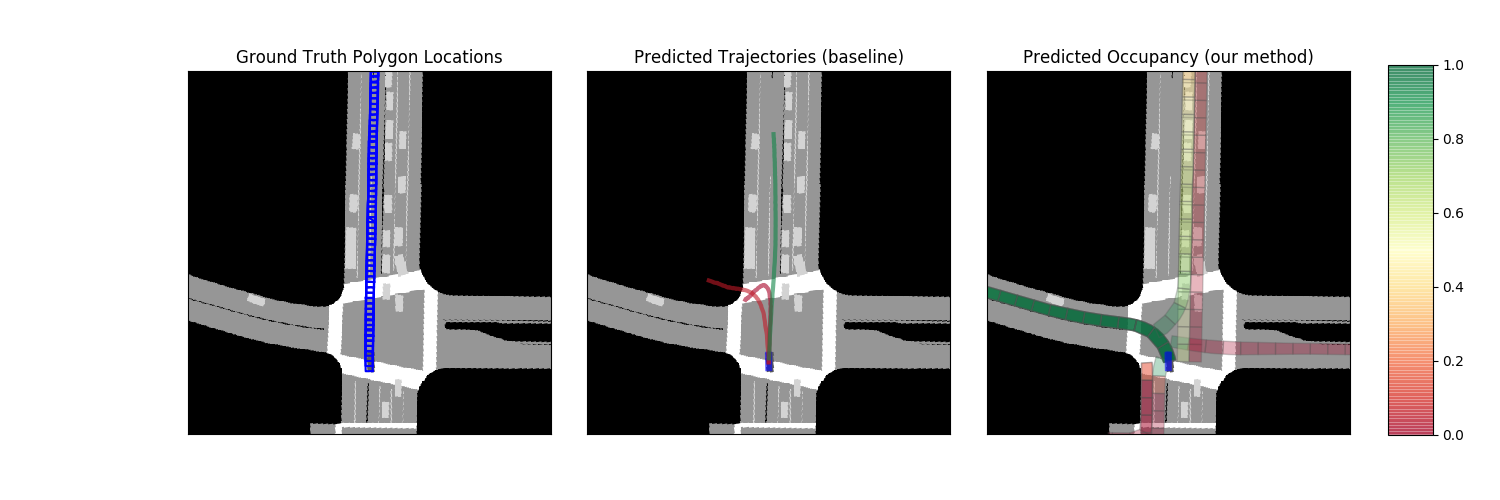}\\[3pt]
\includegraphics[width=\linewidth, trim=4.7cm 1.6cm 3.7cm 1.8cm, clip]{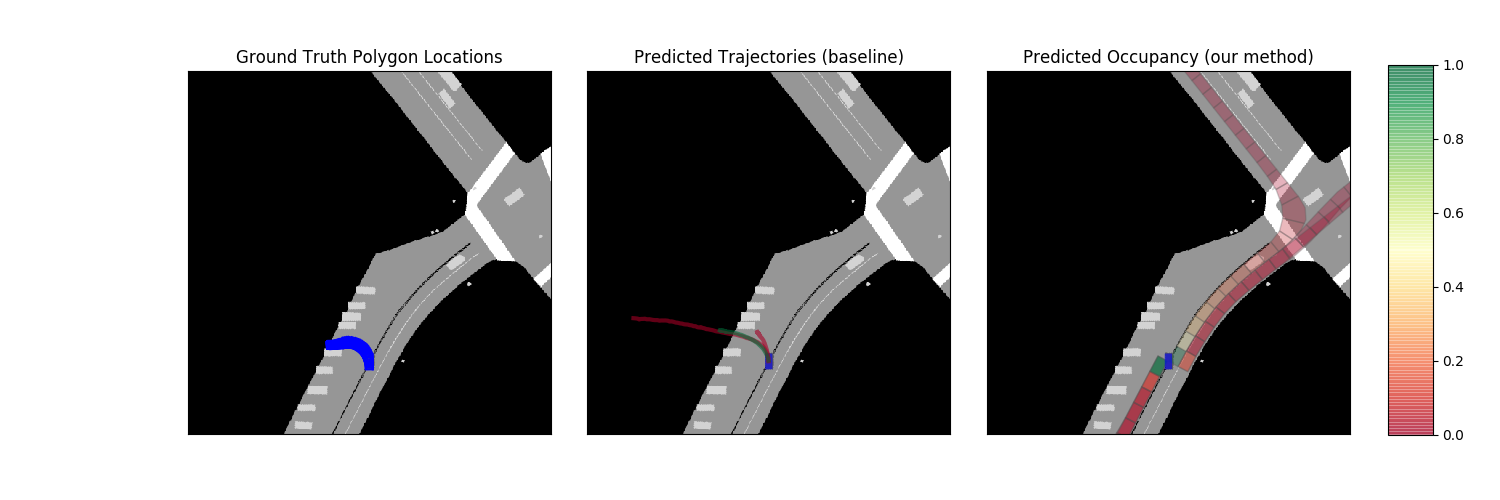}\\[3pt]
\includegraphics[width=\linewidth, trim=4.7cm 1.6cm 3.7cm 1.8cm, clip]{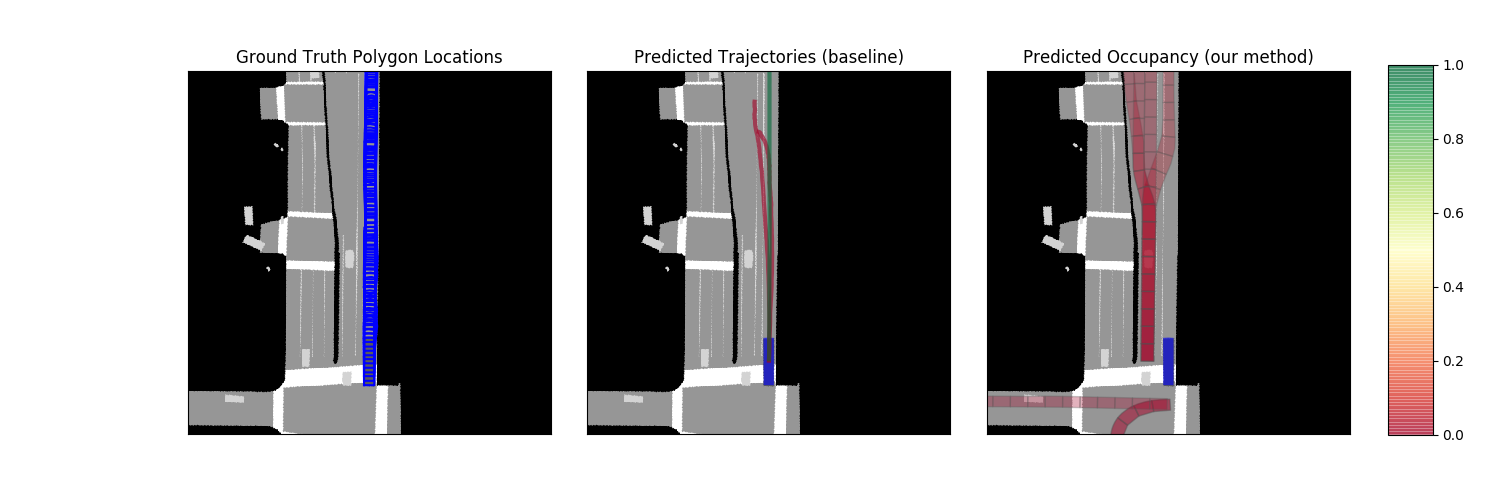}\\[3pt]
\includegraphics[width=\linewidth, trim=4.7cm 1.6cm 3.7cm 1.8cm, clip]{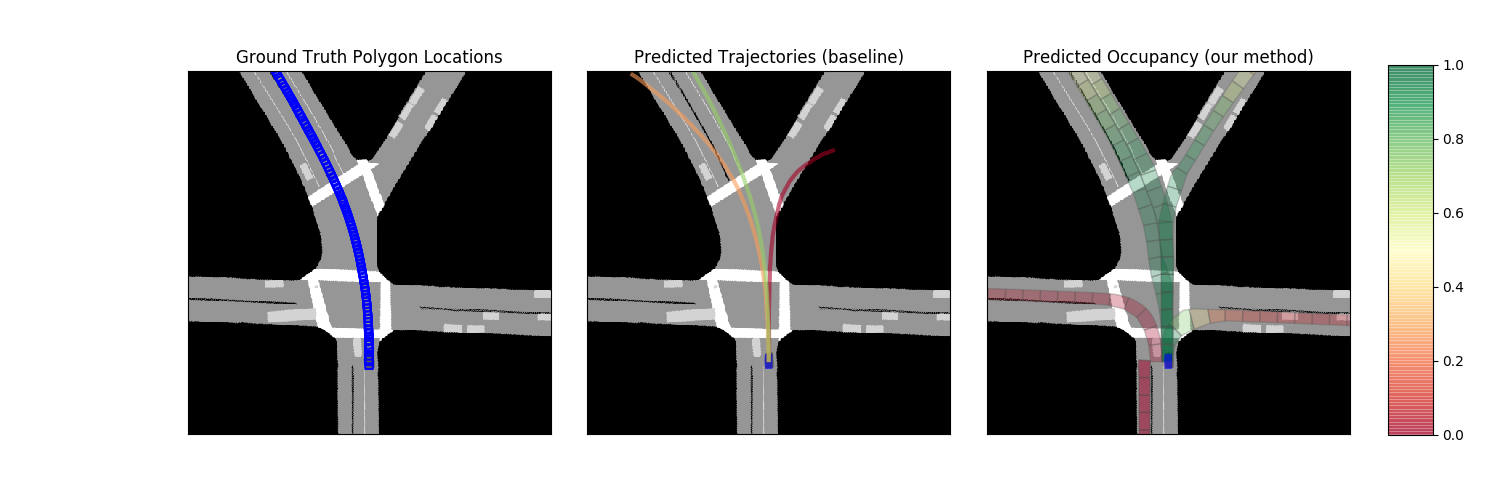}\\[3pt]
{\small (b) Bottom Cases}
\end{minipage}
\caption{Examples of scenes in which our method has (a) higher likelihood and (b) lower likelihood than the MTP baseline. Within each panel, the left image shows the ground truth, the center image shows the trajectory predictions from MTP (each trajectory is colored according to its mode probability), and the right image shows the occupancy predictions from our method (each cell is colored according to its occupancy likelihood). (a) The top two examples depict unusual road geometries (here MTP predicts that the vehicle will drive out of the road or into an oncoming lane). The bottom two examples are cases where the baseline under-utilizes the map information (e.g., knowledge of left-turn-only lanes). (b) The first example shows an emergency vehicle driving against the direction of the lane. The next two examples show actors driving in areas where we don't have mapped lanes like parking spaces and shoulders). The last example shows a case where our predictions are more multimodal than MTP.}
\label{fig:top-and-bottom-cases}
\end{figure}

To further understand our results, we examine the cases with the best and worst likelihood deltas compared to the MTP baseline. Figure~\ref{fig:top-and-bottom-cases} shows examples from the results where our method performed better and worse than the baseline. The top left example shows that MTP rigidly predicts left, right, and straight trajectories relative to the actor's position even when the lane topology doesn't warrant it. The right panel illustrates some cases where we achieve a lower likelihood than MTP, such as when actors drive in unmapped areas. Interestingly, the bottom right example shows a case where we get penalized in the overall likelihood and negative likelihood metrics for having significant but appropriate multi-modality in our predictions, whereas MTP only predicts one reasonable mode.
%some failure modes of our approach, including difficulty predicting non-compliant behaviors such as an actor driving the wrong way in a lane, as shown in the top right example. 

\begin{figure}[!ht]
\centering
\begin{minipage}[t]{0.48\linewidth}
{\begin{flushleft} \small ~~ Truth ~~~~~ MTP ~~~~~  LON   \end{flushleft}}
\vspace{-1mm}
\includegraphics[width=\linewidth, trim=4.7cm 1.6cm 3.7cm 1.8cm, clip]{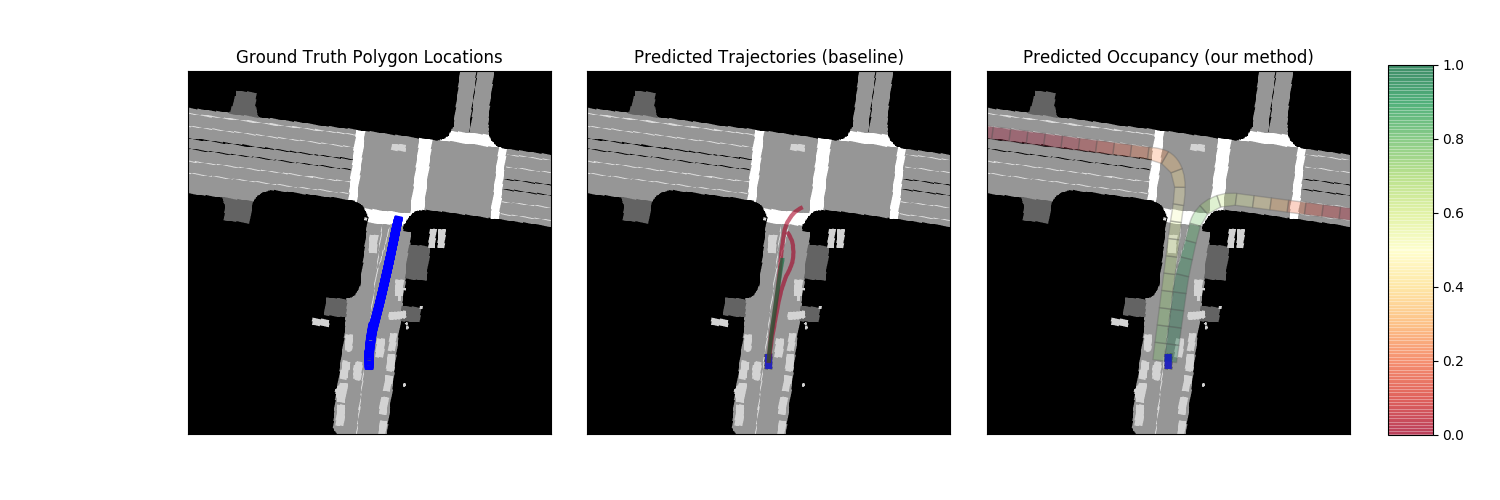}
\end{minipage}
\hfill
\begin{minipage}[t]{0.48\linewidth}
{\begin{flushleft} \small  ~~ Truth ~~~~~ MTP ~~~~~  LON \end{flushleft}}
\vspace{-1mm}
\includegraphics[width=\linewidth, trim=4.7cm 1.6cm 3.7cm 1.8cm, clip]{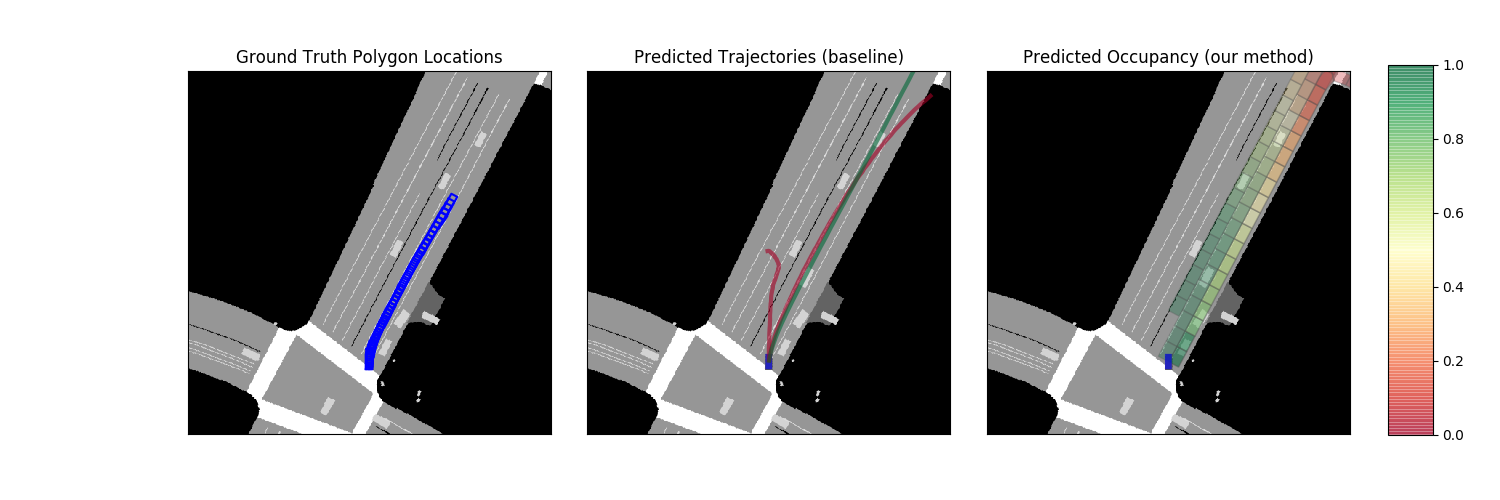}
\end{minipage}
\caption{Examples showing that our method can handle complex scenarios involving actors moving across multiple lanes. The example on the left shows the actor going around a blocking actor and coming back to the lane. The example on the right shows a lane change, where we rightly predict less likelihood for future occupancy in the current lane and more likelihood in the adjacent lanes.}
\vspace{-.3cm}
\label{fig:nudges}
\end{figure}

We also highlight our performance on a few specific cases of interest in which actors move between different lane paths. Figure~\ref{fig:nudges} shows examples where our method correctly assigns high likelihood to adjacent lanes in order to capture an actor going around another actor (left) and a lane change (right). %method predicts occupancy that accurately captures lane changing behavior and go around blocking actors. 

\section{Conclusion}
\label{sec:conclusion}

We present LaneOccupancyNet, a model that incorporates map structure in a novel way to predict the future occupancy of lane-following vehicles. Through quantitative and qualitative results, we demonstrate that our method does a better job than the two baselines of capturing the full distribution over possible future behaviors of an actor. In autonomous driving, the ability to generate multimodal predictions with high recall is critical for safe operation of SDVs. 

By predicting discretized occupancies along a lane path, we find a middle ground between unstructured trajectory predictions and strict lane-following predictions, as demonstrated by our model's ability to predict complex lane-changing maneuvers while still generalizing well to unusual road topologies. Although our approach is sensitive to good map coverage, in that we are only able to predict occupancy for mapped lanes, we assume an SDV must drive strictly within the mapped road network, thus our method still predicts occupancy in regions most important to the SDV. We also show, through qualitative examples, that it can capture a full range of behaviors including those where actors move between lanes rather than simply driving along a single lane.

\addtolength{\textheight}{-11cm}   % This command serves to balance the column lengths

%%%%%%%%%%%%%%%%%%%%%%%%%%%%%%%%%%%%%%%%%%%%%%%%%%%%%%%%%%%%%%%%%%%%%%%%%%%%%%%%

%%%%%%%%%%%%%%%%%%%%%%%%%%%%%%%%%%%%%%%%%%%%%%%%%%%%%%%%%%%%%%%%%%%%%%%%%%%%%%%%



% \section*{ACKNOWLEDGMENT}

% The preferred spelling of the word ÒacknowledgmentÓ in America is without an ÒeÓ after the ÒgÓ. Avoid the stilted expression, ÒOne of us (R. B. G.) thanks . . .Ó  Instead, try ÒR. B. G. thanksÓ. Put sponsor acknowledgments in the unnumbered footnote on the first page.

%%%%%%%%%%%%%%%%%%%%%%%%%%%%%%%%%%%%%%%%%%%%%%%%%%%%%%%%%%%%%%%%%%%%%%%%%%%%%%%%

\bibliography{references}  % .bib
\bibliographystyle{ieeetr}

\end{document}